# Unexpected band gap increase in the Fe$_2$VAl Heusler compound

Alexandre Berche, Martin Talla Noutack, Marie-Liesse Doublet and Philippe Jund*

ICGM, Univ. Montpellier, CNRS, ENSCM, Montpellier, France

Corresponding author: P. Jund
E.mail: philippe.jund@umontpellier.fr
Permanent Address: Université de Montpellier, Place Eugène Bataillon, CC1501, 34095 Montpellier, France

**Abstract**

Knowing the electronic structure of a material is essential in energy applications to rationalize its performance and propose alternatives. Materials for thermoelectric applications are generally small-gap semiconductors and should have a high figure of merit ZT. Even if the Fe$_2$VAl Heusler compound has a decent ZT, its conductive nature (semimetal or semiconductor) is not yet clarified especially at low temperature. In this paper, we focus our DFT calculations on the effect of temperature on the bandgap of Fe$_2$VAl. In contrast to what is usually observed, we show that both the temperature increase and the formation of thermally-activated Al/V inversion defects (observed experimentally), open the bandgap. Such an unusual behavior is the key for reconciling all bandgap measurements performed on the Fe$_2$VAl compound using a standard GGA functional and could be an efficient way for improving the thermoelectric properties of this family of materials.

For thermoelectric applications, the full Heusler Fe$_2$VAl compound may be a good challenger to conventional bismuth telluride (Bi$_2$Te$_3$) currently used in commercialized thermoelectric refrigerators. Most of the thermoelectric materials are small-gap semiconductors doped to obtain optimal carrier concentrations (around $10^{19}$ charges.cm$^{-3}$) which generally allow to have materials with a high figure-of-merit (named ZT = S$^2\sigma$T/$\kappa$ where T is the temperature, S the Seebeck coefficient, $\sigma$ and $\kappa$ the electrical and thermal conductivity respectively). Fe$_2$VAl exhibits a ZT max around 0.3 at 500K for Tantalum-doped bulk materials [1]. Even if this value is below the target value of ZT = 1, it can probably still be improved since very recently Hinterleitner et al. have shown that Tungsten doped thin films exhibit ZT values as large as 6 at 400K [2]. Nevertheless, despite the intense work dedicated to Fe$_2$VAl based materials, the value and the nature of the band gap of the pure compound is still not yet completely known. Indeed, experimentally, two different behaviors have been reported depending on the measurement method and/or on the temperature of the measurement. At low temperature, photoemission measurements at 40K [3] show the absence of an energy gap at the Fermi level. RMN analyses between 4K and 550K [4] suggest the presence of a gap (0.2-0.3eV) but with a small amount of density of states (DOS) at the Fermi level. Optical conductivity measurements performed between 9K and 295K [5] confirm that Fe$_2$VAl is a semimetal with a pseudo-gap of 0.1eV to 0.2eV. These results at low temperature are consistent with DFT calculations (within the LDA or GGA) whose calculated band structures also show a pseudo-gap with the same magnitude. Nevertheless in these calculations an indirect bandgap E$_g \approx$ -0.1eV (E$_g$ is defined as the energy difference between the bottom of the conduction band and the top of the valence band) is predicted between the high symmetry points $\Gamma$ and X in the first Brillouin zone [6-11]. According to Do *et al.* [7], the residual density of states (DOS) at the Fermi level is due to the presence of d-states of vanadium constituting the bottom of the conduction band while the valence band maximum is due to d-states of iron.

However, such a semimetallic behavior is in contradiction with the activated mechanism reported at higher temperature (300-800K) and observed in electrical measurements, which is typical of a semiconductor compound with a gap around 0.1eV [3, 12]. An explanation of this behavior has been suggested by Okamura *et al.* [5]: $Fe_2VAl$ is a semimetal with a small DOS at the Fermi level. At high temperature, $k_BT$ is comparable to the value of the gap, so that a large number of carriers are thermally excited across the pseudogap explaining the typical « semiconducting » behavior observed at high temperature.

In addition to these band structure studies, Rietveld analyses of neutron diffraction patterns performed on $Fe_2VAl$ [13] have shown that, for T > 1023K (i.e. below the $L2_1$/B2 transition temperature (1353K)), partial disorder is observed due to inversions between Al and V atoms. In the spirit of our previous works on point defects in thermoelectric materials such as NiTiSn [14-16], NiHfSn [17], NiZrSn [18] and ZnSb [19-20], we here investigate the impact of temperature and partial disorder in the stoichiometric $Fe_2VAl$ through first-principles DFT calculations. The effect of off-stoichiometric defects has been discussed in a previous work [11].

Calculations were performed using the Vienna *ab initio* simulation package (VASP) [21-22] and the projector augmented waves (PAW) technique [23-24] within the generalized gradient approximation (GGA). The Perdew-Burke-Ernzerhof parameterization (PBE) was applied [25-26] and standard versions of the PAW potentials for Fe ($3p^63d^64s^2$), V ($3s^23p^63d^34s^2$) and Al ($3s^23p^1$) were used. The first Brillouin zone was sampled using 3x3x3 Monkhorst-Pack k-point meshes [27]. The cutoff energy was set to 500 eV for the whole study. Both cell parameters and atomic positions were relaxed within an energy accuracy of 1 µeV and 10 µeV/Å for the forces.

A 3x3x3 (108 atoms) and a 4x4x4 (256 atoms) supercell of the primitive cell (named P333 and P444 respectively) were used to avoid finite-size effects in the defective phases. Two different inversions were investigated in this study. The first one assumes that the two antisites are nearest neighbors (correlated defects) and the second one considers two randomly chosen sites (uncorrelated defects). For the latter, the Special Quasi-Random Structures (SQS [28]) available in the ATAT software suite [29] was used to avoid artificial ordering effects in the distribution of the atoms. In this methodology, 15 cells have been initially generated and the "less ordered" structure is selected for the DFT calculations.

For all these disordered cells, the $Fe_2VAl$ compound remains stoichiometric. The formation energy of the defect ($\Delta_{def}E$) was then calculated from the equation $\Delta_{def}E = (E(disorder)-E(order))/x_{defect}$ where E(disorder) and E(order) are the total energy of the disordered and the perfect cell respectively (in eV/atom) and $x_{defect}$ is the concentration of antisite defects in the cell (each Al/V inversion corresponding to two antisites).

The thermally activated behavior of the pure $Fe_2VAl$ compound is investigated through *Ab Initio* Molecular Dynamic calculations (AIMD) as implemented in the VASP package. The Verlet algorithm is used to numerically integrate Newton's equations of motion within the Born-Oppenheimer approximation. All AIMD calculations were performed within the (NVT) canonical ensemble i.e. the number of particles N, the volume V, and the temperature T are kept fixed (the temperature is controlled using a Langevin thermostat). All simulations were performed over 8 ps with a time step of 1fs (initial (NVE) runs confirm that the energy is conserved with such a timestep) in the P333 supercell and performed at the Γ point of the Brillouin zone. The convergence criterion over the self-consistent cycles was set at $10^{-6}$ eV. For each temperature, the Fermi-Dirac smearing factor was considered and the equilibrium volume was first determined. For this step, two AIMD runs are performed at two fixed volumes $V_1$ and $V_2$ with $V_1-V_2 \sim 0.3$ Å$^3$ and the equilibrium volume $V_0$ (corresponding to zero pressure) is obtained through a linear interpolation between $(V_1,P_1)$ and $(V_2,P_2)$, where $P_i$ is the corresponding pressure of the system (ideally one of the $P_i$s is negative, the other positive). The system at $V_0$ is relaxed during 5ps in order to stabilize the pressure and the total energy (as can be seen in Supplementary Data A, this equilibration time is more than sufficient) and the equilibrium atomic positions are saved during the last 3ps at each temperature. The subsequent DFT band structure calculations were then performed on ten different equilibrium configurations within a pressure range of +/-0.02 kbar at a given temperature (we have checked that in this range, $E_g$ is not pressure-dependent).

Let us begin with the influence of the defects seen experimentally on the electronic properties of stoichiometric $Fe_2VAl$. To this aim, we have calculated the formation energy of the three possible inversions: Al/V, Al/Fe and Fe/V. For each defect, the two possible configurations (correlated and uncorrelated) have been tested (Table 1). It appears clearly that correlated defects are more probable than uncorrelated ones. Moreover, the correlated Al/V inversion is clearly the most favorable defect in stoichiometric $Fe_2VAl$. This result is in agreement with the experimental results of Maier *et al.* [13] and is also consistent with the crystal structure of the Heusler phase (space group Fm-3m, n°225) in which V (*4a* positions (0;0;0)) and Al (*4b* positions (½;½;½)) have the same environment (8 Fe atoms as first nearest neighbors) while Fe atoms (*8c* positions (¼;¼;¼)) have 4 Al and 4 V atoms as first neighbors in a stellated tetrahedron configuration. As shown in Figure 1b, this Al/V inversion significantly affects the band structure with $E_g$ increasing up to a value of -0.006eV and the two bands crossing the Fermi level being less dispersive. Larger defect concentrations (corresponding to a higher temperature in experiments) were thus investigated to check the tendency of these specific Al/V inversions to open a bandgap in the stoichiometric $Fe_2VAl$ compound.

Table 1: Formation energy of defects for correlated and uncorrelated inversions.

| Inversion | $\Delta_{def}E$ (eV.atom$^{-1}$) | |
| --- | --- | --- |
| | Correlated | Uncorrelated |
| Al/V | 0.342 | 0.405 |
| Fe/V | 1.278 | 1.674 |
| Al/Fe | 1.490 | 1.730 |

For the second Al/V inversion, ten configurations have been tested. The second inversion is the most probable ($\Delta_{def}E$ = 0.277 eV.atom$^{-1}$) when it is located close to the first one, in a square like configuration (Figure 2b). By comparison, the presence of a second inversion uncorrelated to the first one is less favorable ($\Delta_{def}E$ = 0.304 eV.atom$^{-1}$). In this most stable configuration, two iron atoms (nearest neighbors of the inversions) loose two of their Fe-V bonds. A look at the band structure close to the Fermi level indicates that the loss of these Fe-V bonds could have significant impact on the value of the band gap. Indeed, with two Al/V inversions, a positive $E_g$ of +0.078eV is predicted (see Figure 1c). Thus our DFT calculations suggest that $Fe_2VAl$ which is a semimetal at 0K becomes semiconductor at higher temperatures due to an increasing concentration of Al/V inversions. This mechanism is an alternative explanation to the one proposed by Okamura *et al.* [5] who claim that $Fe_2VAl$ is a semimetal at low temperature with such a small density of states at the Fermi level that at high temperature a large number of carriers are thermally excited across the pseudogap leading to the typical « semiconducting » behavior observed at high temperature.

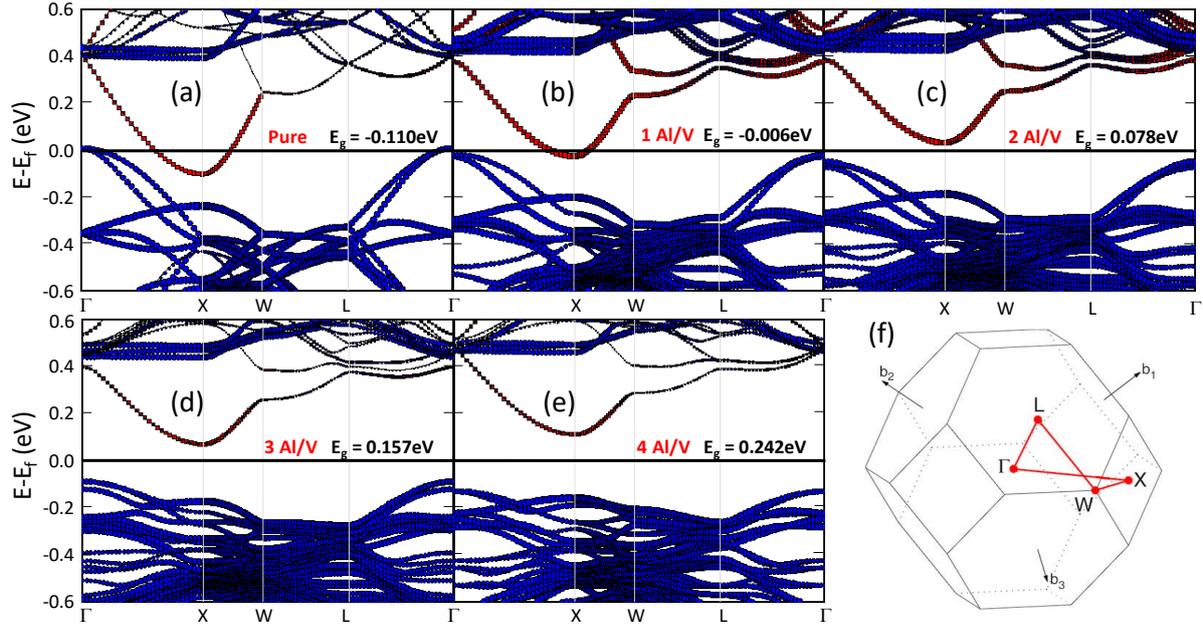

Figure 1: Band structure of a) pure Fe$_2$VAl and b) 1Al/V, c) 2Al/V, d) 3Al/V and e) 4Al/V inversions in a P333 supercell (first Brillouin zone is given in f)). Red squares represent the d-states of V while blue circles show the d-states of Fe.

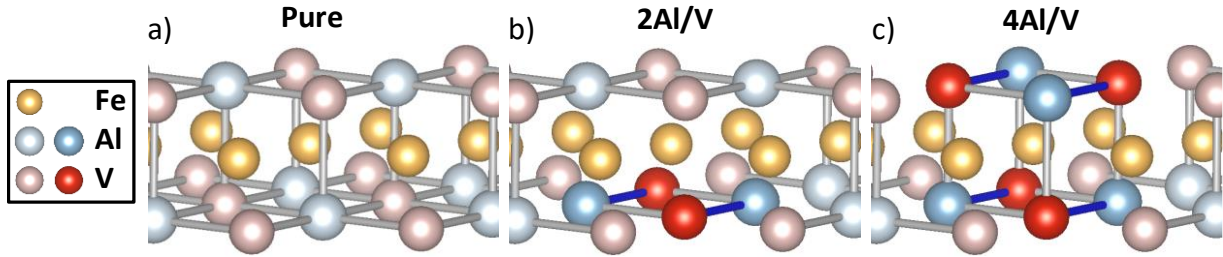

Figure 2: Enlargement of the crystal structure around inversions in the most probable configuration: a) pure compound; b) two Al/V inversions; c) four Al/V inversions. Atoms in antisite positions are in bright blue (Al) and red (V) and are linked by a blue bond.

When a third Al/V inversions is added to the supercell, the most probable configuration ($\Delta_{def}E = 0.246$ eV.atom$^{-1}$) is obtained when the inversion occurs directly above (or below) the previous square but in the contiguous Al-V plane. Other configurations have been tested. If the 3$^{rd}$ inversion is in the plane of the previous square ($\Delta_{def}E = 0.261$ eV.atom$^{-1}$) or if the 3$^{rd}$ inversion is far from the first two inversions ($\Delta_{def}E = 0.294$ eV.atom$^{-1}$), the configurations are less probable. The most stable configuration with three Al/V inversions prefigures what happens when a 4$^{th}$ inversion occurs (Figure 2 c). Indeed, the 4$^{th}$ Al/V inversion is most stable in a configuration where a cubic antiphase cluster is built around one Fe atom (all its neighbors have been exchanged). The addition of a third and a fourth Al/V inversion increases the value of $E_g$ to +0.157eV (Figure 1d) and +0.242eV, respectively (Figure 1e). It is worth noting that the value of the calculated bandgap varies linearly (Figure 3) with the fraction of Al (or V) sites occupied by V (or Al) atoms, for substitution rates ranging from 0 to 15%. This behavior has been checked in a P444 supercell to have an intermediate substitution rate. As shown in Figure 3, the linear variation of $E_g$ with the fraction of Al/V substitutions is not cell size dependent.

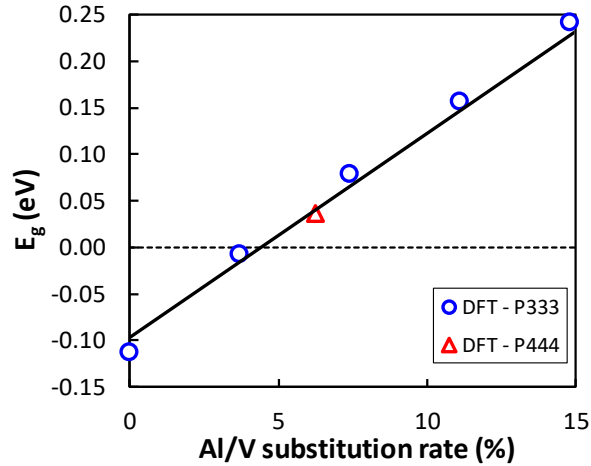

Figure 3: Evolution of the value of $E_g$ with the Al/V substitution rate.

Based on the formation energy of the different defects ($\Delta_{def}E$), the evolution of the substitution rate with temperature was then calculated from $\exp(-\Delta_{def}E/(kT))$. For a single Al/V inversion, $\Delta_{def}E = 0.342$ eV.atom$^{-1}$ is used and this formula is only correct if the substitution rate is between 0 and 1/27 (rate corresponding to 1 inversion). Above this value, the formation energy of 2Al/V inversions has to be used up to a substitution rate of 2/27, and so on. The obtained evolution of the Al/V substitution rate (Figure 4) is in perfect agreement with the measurements of Maier *et al.* [13], which validates our approach.

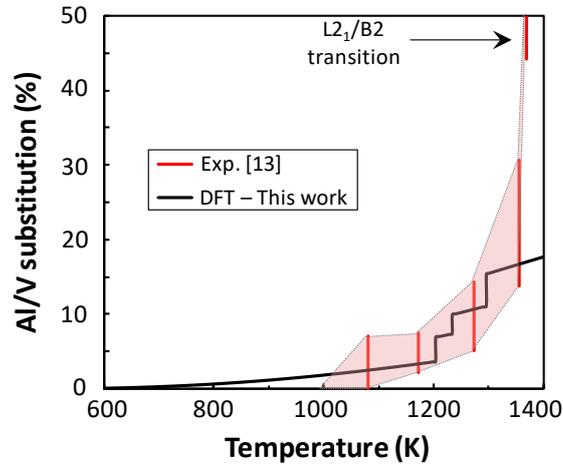

Figure 4: Evolution of the Al/V substitution rate as a function of temperature compared to experiments.

Even though the presence of Al/V inversions increases the value of $E_g$, it is probable that almost no inversion exists below 900K-1000K at thermodynamic equilibrium (Figure 4). These thermally-activated defects can therefore not explain the semiconducting behavior reported at 400K - 800K. To account for the effect of temperature on the band structure (and therefore on $E_g$) AIMD calculations were performed on pure $Fe_2VAl$ at 3 temperatures: 300K, 600K and 900K. At thermodynamic equilibrium, AIMD predicts the thermal expansion of the compound together with positional disorder leading to enhanced electron-phonon interactions. Both phenomena increase $E_g$ when the temperature increases as shown in Figure 5 (see also Supplementary Data B for typical band structures of finite temperature structural snapshots). Such an abnormal behavior has only been reported for the PbX- type family of compounds (X = S, Se, Te) both in simulations similar to the present ones and in experiments [30]. In Figure 5, the contribution of the thermal expansion to the bandgap increase is compared to the more global thermal contribution including electron-phonon interactions. As can be seen, in $Fe_2VAl$ the thermal expansion only contributes for approximately 20% to the global bandgap increase (comparison between black continuous line and dotted black line in Figure 5) as opposed to almost 50% in the PbX compounds [30].

Thus positional disorder due to thermal vibrations and thermally-activated Al/V inversions (Figure 3) are the main contributors to the bandgap increase with temperature in $Fe_2VAl$. Both these contributions lead to a break of symmetry, as opposed to the thermal expansion, and it is therefore in that direction that the fundamental origin of the bandgap increase should be looked for rather than in a simple bond length increase phenomenon.

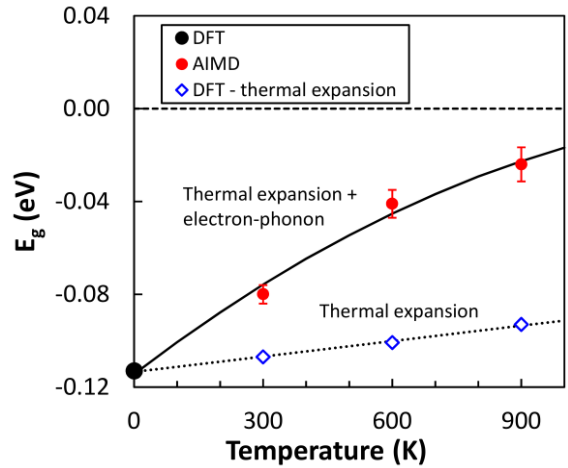

Figure 5: Evolution of the calculated $E_g$ (DFT and AIMD) as a function of temperature. The continuous black line represents the general tendency of the evolution due to the thermal expansion and to the thermal disorder while the dotted line only represents the effect of the thermal expansion. The contribution of thermal expansion was determined in $Fe_2VAl$ cells with the calculated expanded cell parameters obtained from the AIMD simulations.

To conclude, with DFT calculations at 0K on defective cells or with AIMD at finite temperatures on pure cells, we have shown that $Fe_2VAl$ is a new material exhibiting an unusual temperature-dependent bandgap. When temperature increases, three phenomena occur in $Fe_2VAl$: thermal expansion, thermal disorder and the formation of thermally-activated Al/V inversions that appear significantly above 800K. In experimental samples, these mechanisms combine and open the bandgap when temperature increases, which explains the difference in the experimental observations between optical measurements at low temperature and electrical measurements at high temperature. It also solves the dispute between different DFT calculations trying to find a semiconductor at 0K with the use of elaborated (and computer time costly) functionals [6-8] in order to be in agreement with experiments at high temperature. We show that taking into account temperature effects via AIMD simulations and/or Al/V inversion defects in the stoichiometric $Fe_2VAl$ compound through a supercell approach, solves this apparent disagreement using standard GGA calculations. It is worth noting that if the GGA allows to reproduce the bandgap increase due to temperature, the absolute value of $E_g$ is likely underestimated which is a regular propensity of the GGA functionals. Of course, other functionals such as mBJ [31-32] or SCAN [33] can be used but the ground state at 0K will be calculated as a semiconductor (see Supplementary Data C) which is not consistent with experimental data collected at low temperature, measuring a non-zero density of states at the Fermi level.

Nevertheless, irrespectively of the functional used in the calculations, Al/V inversions invariably open the bandgap of $Fe_2VAl$ (see Supplementary Data C) showing that this result is not methodology-dependent but has a physical and/or chemical origin.

The increase of the bandgap with temperature shown here is favorable for thermoelectric materials since it can reduce the bipolar effects caused by the activation of the minority charge carriers and thus lead to a higher ZT at high temperatures. In addition, the defect influenced electronic behavior shown in $Fe_2VAl$ can also provide a new route to obtain $Fe_2VAl$ based thermoelectric materials with high figures-of-merit. Indeed, if it were possible to fix the high temperature disordered state (with an ultra-quench from a 1300K annealed material for example), such a material would have a significantly higher band gap which will increase the value of the Seebeck coefficient, and as a consequence the Power Factor (PF= $S^2\sigma$) of the sample. Moreover, the presence of a significant amount of antiphases will generate interfaces between the grains of $Fe_2VAl$ resulting in a decrease of the thermal conductivity $\kappa$ which is also beneficial for increasing the figure-of-merit of the compound.

**Credit author statement:**
Alexandre Berche : Methodology ; Investigation ; Data curation ; Writing original draft ; Visualization
Martin Talla-Noutack :  Methodology ; Investigation ; Data curation
Marie-Liesse Doublet : Writing – Review & Editing ;  Validation ; Formal analysis
Philippe Jund : Conceptualization; Resources; Writing - Review & Editing; Supervision; Project administration;  Funding acquisition

**Acknowledgments:** The financial means to perform this work have been provided by the French Science Foundation (ANR project LoCoThermH). We thank E. Alleno and J.M. Joubert from ICMPE (Thiais, France) for helpful discussions.

# Supplementary data

**A- Evolution of pressure and energy during the AIMD run at equilibrium volume (900K)**

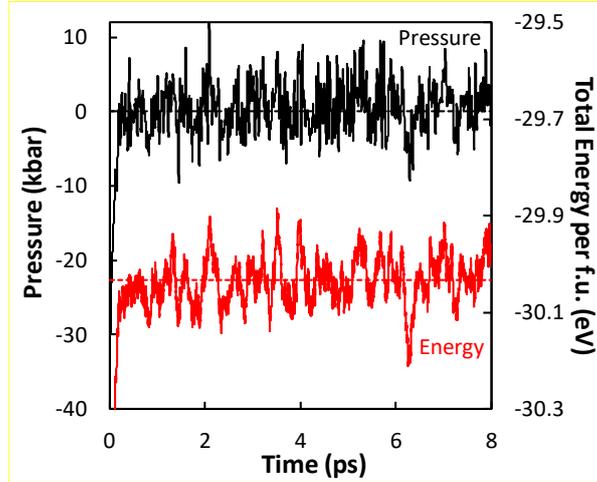

Figure A: Evolution of the pressure and the total energy of the system during the AIMD calculation at 900K

**B- Effect of temperature on the band structure of $Fe_2VAl$ (AIMD)**

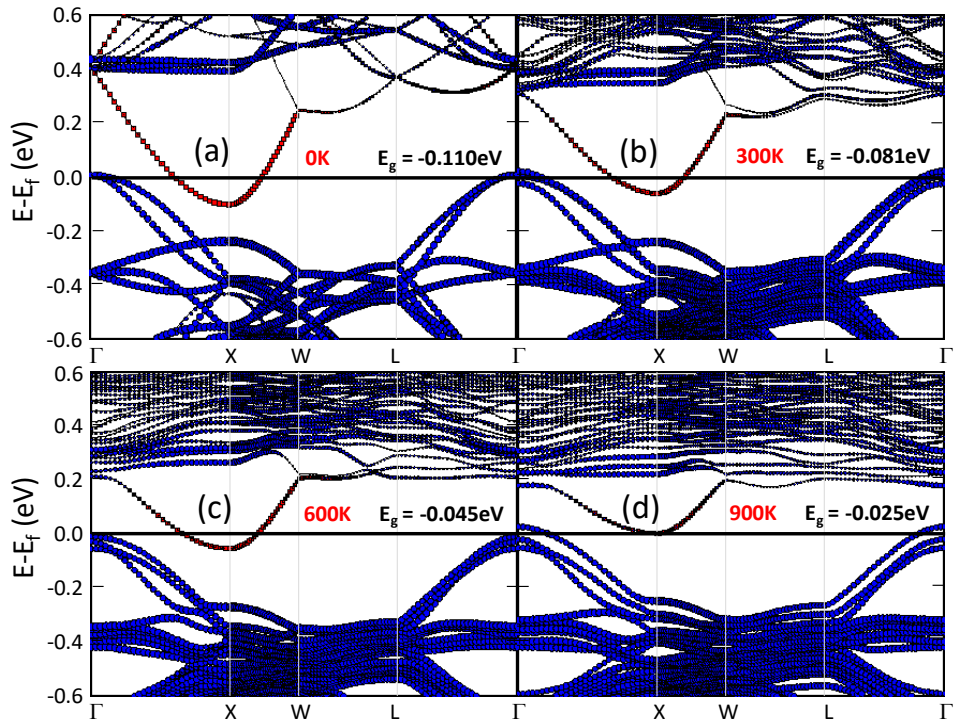

Figure B: Effect of the temperature on the band structure of pure $Fe_2VAl$ according to *ab initio* calculations at: a) 0K (DFT); b) 300K (AIMD); c) 600K (AIMD) and d) 900K (AIMD)

## C- Bandgap increase using different DFT functionals

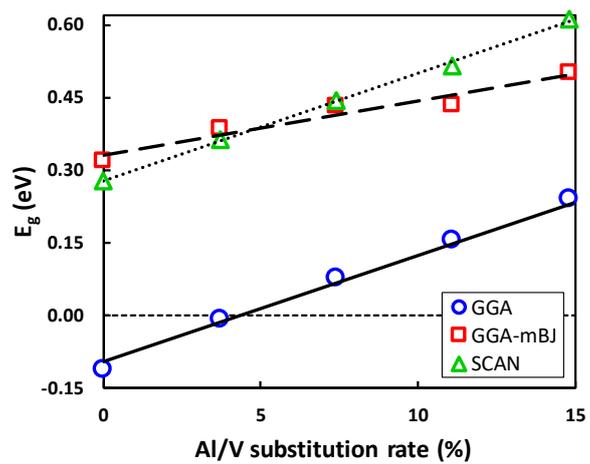

Figure C: Effect of the choice of the functional (GGA, GGA+mBJ or SCAN) on the increase of $E_g$ vs the Al/V substitution rate.